\documentclass[12pt]{article}
\pdfoutput=1
\usepackage[margin=3cm]{geometry}
\usepackage[utf8]{inputenc}
\usepackage{amsmath}
\usepackage{amssymb}
\usepackage{authblk} % for authors
\usepackage{xcolor}
\usepackage{physics}
\usepackage{graphicx} % for graphics in pdf_tex figures
\usepackage{float}
\usepackage{tikz}
\usetikzlibrary{calc}
\usepackage{comment}

\usepackage{hyperref}
\usepackage{varioref}       
\usepackage{cleveref}     

\crefname{table}{table}{tables}
\Crefname{table}{Table}{Tables}
\crefname{figure}{figure}{figures}
\Crefname{figure}{Figure}{Figures}

\definecolor{tealblue}{rgb}{0.21, 0.56, 0.63}
\hypersetup{colorlinks=true,allcolors = tealblue,linktocpage=true}

\usepackage{pgf} % for pgf pics, might have to remove on publication if pgfs have to be prerendered

\usepackage[
        backend=bibtex,
        sorting=none,
        style=phys,
        eprint=true,
        doi=false,
        biblabel=brackets
        ]{biblatex}

\bibliography{dsdc}

\usepackage[backgroundcolor=white!95!red, linecolor=cyan, bordercolor=cyan, textsize=footnotesize]{todonotes}

\newcommand{\commie}[1]{}

\numberwithin{equation}{section}
\numberwithin{table}{section}

\newenvironment{eqaed}
    {\begin{equation}
    \begin{aligned}
    }
    { 
    \end{aligned}
    \end{equation}
    \ignorespacesafterend
    }

\begin{document}

\title{Domain walls and distances in discrete landscapes}
\date{}

%\author{Ivano Basile$^{a}$\thanks{ivano.basile@lmu.de}$\;$, $\;$ Carmine Montella$^{b}$}
%\affil{\emph{$^a${\it Arnold Sommerfeld Center for Theoretical Physics,\\
%Ludwig-Maximilians-Universit\"at M\"unchen, 80333 M\"unchen, Germany}}\\[0.1cm]
%\vspace{2mm}
%\emph{$^b${\it Max-Planck-Institut f\"ur Physik (Werner-Heisenberg-Institut), \\
%   F\"ohringer Ring 6,  80805 M\"unchen, Germany}}}

\author{Ivano Basile\thanks{ivano.basile@lmu.de}}
\affil{\emph{Arnold-Sommerfeld Center for Theoretical Physics}\\ \emph{Ludwig Maximilians Universit\"at M\"unchen}\\ \emph{Theresienstraße 37, 80333 M\"unchen, Germany}}

\author{Carmine Montella\thanks{montella@mpp.mpg.de}}
\affil{\emph{Max-Planck-Institut f\"ur Physik (Werner-Heisenberg-Institut)}\\ \emph{F\"ohringer Ring 6,  80805 M\"unchen, Germany}}

\maketitle

\begin{abstract}
    
    We explore a notion of distance between vacua of a discrete landscape that takes into account scalar potentials and fluxes via transitions mediated by domain walls. Such settings commonly arise in supergravity and string compactifications with stabilized moduli. We derive general bounds and simple estimates in supergravity which constrain deviations from the ordinary swampland distance conjecture based on moduli space geodesics, and we connect this picture to renormalization group flows via holography.
    
\end{abstract}

\thispagestyle{empty}

\newpage

\tableofcontents

\thispagestyle{empty}

\newpage

\pagenumbering{arabic}

\section{Introduction}\label{sec:introduction}

Among various ideas that have emerged in the context of the swampland program~\cite{Vafa:2005ui} (see~\cite{Brennan:2017rbf, Palti:2019pca, vanBeest:2021lhn, Grana:2021zvf, Agmon:2022thq} for reviews), the ones related to dualities and infinite-distance limits~\cite{Ooguri:2006in, Baume:2016psm, Klaewer:2016kiy, Blumenhagen:2018nts, Heidenreich:2018kpg, Baume:2020dqd, Perlmutter:2020buo, Grimm:2022sbl, Etheredge:2022opl, Baume:2023msm, Etheredge:2023odp, Calderon-Infante:2023ler, Cribiori:2023ffn} stand out for a number of reasons. To begin with, they pinpoint a specific way in which an effective field theory (EFT) breaks down due to quantum gravity effects, namely due to a tower of \emph{infinitely many} species that becomes light with respect to the relevant Planck scale. Constraints for and due to this behavior can be quantified more precisely relative to more kinematical aspects of quantum gravity, such as the absence of global symmetries. Furthermore, the emergence of infinite towers at low energies can drastically modify the physics, constraining phenomenology (see \emph{e.g.}~\cite{Scalisi:2018eaz, Scalisi:2019gfv, Montero:2022prj, Anchordoqui:2023oqm}) and typically leading to a new duality frame controlled by different light degrees of freedom expected to be higher-dimensional quantum fields or weakly coupled strings~\cite{Lee:2018urn, Lee:2019xtm, Lee:2019wij, Lanza:2020qmt, Lanza:2021udy, Lanza:2022zyg}. To be more precise, the existence of light towers and the specific asymptotics of their mass scales can be considered separately, and the more quantitative statement concerns the second. A generalized version of the conjecture can be summarized by the following statement: along a divergent sequence $\{p_n\}$ of points in the set of vacua of some EFT consistently coupled to quantum gravity in $d$ spacetime dimensions, the mass scale of the (lighest) tower scales according to
    \begin{eqaed}\label{eq:SDC}
        \frac{m(p_n)}{m(p_0)} \overset{n \to \infty}{\sim} k \, e^{- \lambda \, \Delta(p_n,p_0)/M_\text{pl}^{\frac{d-2}{2}}} \, ,
    \end{eqaed}
where $\Delta$ is a suitable distance. The positive $\mathcal{O}(1)$ constants $k \, , \, \lambda$ may be bounded in some settings~\cite{Perlmutter:2020buo, Etheredge:2022opl, Calderon-Infante:2023ler}, but it is important to emphasize that the peculiar exponential falloff seems to be the feature that is most closely related to string theory~\cite{Basile:2022sda}.

The aim of this paper is to explore this idea in settings where there is no exact moduli space, where $\Delta$ can be taken as the geodesic distance according to the canonical Riemannian metric. This is relevant when scalar potentials or fluxes are present, for instance in string compactifications with stabilized moduli, and establishing a generalized distance conjecture may shed light on the status of debated proposals for parametrically scale-separated stabilized vacua (see \emph{e.g.}~\cite{Shiu:2022oti} on the DGKT vacuum~\cite{DeWolfe:2005uu}, and also~\cite{Carrasco:2023hta}). While this idea has been considered in various contexts~\cite{Buratti:2018xjt, Lust:2019zwm, Calderon-Infante:2020dhm, Basile:2020mpt, DeBiasio:2020xkv, Stout:2021ubb, Basile:2022zee, Stout:2022phm, Basile:2022sda, Shiu:2022oti, Li:2023gtt}, there seems to be no systematic agreed-upon way to approach the issue. Since a moduli space is a space of vacua parametrized by the expectation values of certain fields or operators, when the set of vacua is discrete our approach is to take it seriously and attempt to define a suitable notion of distance between the vacua, taking into account any ingredient that lifts the moduli space into such a discrete landscape. In order to do this, we base our construction on domain walls. If the EFT at stake contains light scalar fields $\{ \phi^i \}$ whose moduli space with metric $G$ is lifted to a discrete set of vacua $\{ \phi^i_0(\mathbf{n}) \}_{\mathbf{n} \in \Gamma}$ indexed by a lattice $\Gamma$, domain walls are finite-energy solutions that interpolate between two isolated vacua. The important point is that the trajectory in field space induced by the domain wall takes into account scalar potentials and/or fluxes, so that the length of this (non-geodesic) trajectory (as defined by the metric $G$) may be interpreted as a discrete distance between two vacua in a discrete landscape. This notion has a natural holographic counterpart in the form of renormalization group (RG) flows, whose length can be computed using the (quantum) information metric~\cite{oconnor, Stout:2021ubb, Basile:2022zee, Stout:2022phm, Basile:2022sda}. Before outlining the remainder of this paper, let us emphasize that the existence of domain walls (or vacuum bubbles) is expected to be generic on the grounds of cobordism triviality~\cite{McNamara:2019rup}, which means that this theoretical basis for discrete distances can extend beyond the controlled settings of Bogomol'nyi-Prasad-Sommerfield (BPS) domain walls in supergravity and holography.

With this construction at hand, we are naturally led to a discrete version of the distance conjecture. Starting with a theory that satisfies the ordinary incarnation on its moduli space, after lifting it to a discrete set of isolated vacua one can ask whether the towers of species still become light exponentially fast in the discrete distance induced by domain wall transitions. Intriguingly, this criterion can be verified entirely within the landscape of the EFT, without reference to any tower of species or ultraviolet (UV) completion. We shall see how this proposal is realized in string compactifications and how it may fail in other cases.

The contents of this paper are organized as follows. In~\cref{sec:DW_distance} we present the precise proposal on how to define distances between isolated vacua in a discrete landscape. In~\cref{sec:DW_bounds} we derive some bounds and some more qualitative estimates that constrain the behavior of light towers with respect to the discrete distance for BPS domain walls in supergravity. In~\cref{sec:examples} we discuss some simple examples, before moving to the holographic counterpart of this construction in~\cref{sec:holography}. We conclude in~\cref{sec:conclusions} with some final remarks.

\section{Distances in discrete vacua from domain walls}\label{sec:DW_distance}

For the majority of this paper, the general setting we consider is an EFT with some light scalar fields $\{ \phi^i \}$ described by the kinetic term
\begin{eqaed}\label{eq:scalar_kinetic_term}
    \mathcal{L}_\text{kin} = - \, \frac{1}{2} \, G_{ij}(\phi) \, \partial \phi^i \cdot \partial \phi^j \, .
\end{eqaed}
The Riemannian metric $G$ is defined on the space of field values, but we consider cases in which the latter is not a moduli space. Rather, the set of vacua is discrete, lifted by a scalar potential and/or fluxes which we do not need specify yet. We assume that the set of field values takes the form $\{ \phi^i_0(\mathbf{n}) \}_{\mathbf{n} \in \Gamma}$, a set indexed by a lattice $\Gamma$. We denote the corresponding vacua by $|\mathbf{n}\rangle$, even though most of our discussion until~\cref{sec:holography} is going to be (semi)classical.

In specific examples, most notably in the controlled context of minimal supergravity in four dimensions~\cite{Cvetic:1992bf, Cvetic:1992st, Cvetic:1992sf, Cvetic:1993xe, Cvetic:1996vr, Ceresole:2006iq, Bandos:2018gjp, Bandos:2019wgy}, domain walls can be found as codimension-one solitons interpolating between two vacua
\begin{eqaed}\label{eq:DW_interpolation}
    \phi^i_0(\mathbf{n}_1) \quad \overset{\text{DW}_{\mathbf{n}_1 \to \mathbf{n}_2}}{\longrightarrow} \quad \phi^i_0(\mathbf{n}_2)
\end{eqaed}
along a transverse coordinate $r$, defining a profile $\phi_\text{DW}(r)$ for the scalars (we suppress the index whenever possible). Singling this coordinate out, the spacetime line element induced by the backreaction of the domain wall, written in terms of $r$ and longitudinal coordinates $x$, takes the form
\begin{eqaed}\label{eq:DW_metric}
    ds^2_\text{DW} = dr^2 + e^{2A(r)} \, dx^2 \, ,
\end{eqaed}
where $dx^2$ denotes the flat line element. Here we consider domain walls in the strict sense, flat and stationary. However, similar considerations can be made on vacuum bubbles~\cite{Basile:2022zee} which typically arise in metastable non-supersymmetric settings.

\begin{figure}[ht!]
    \centering
    \includegraphics[scale=0.5]{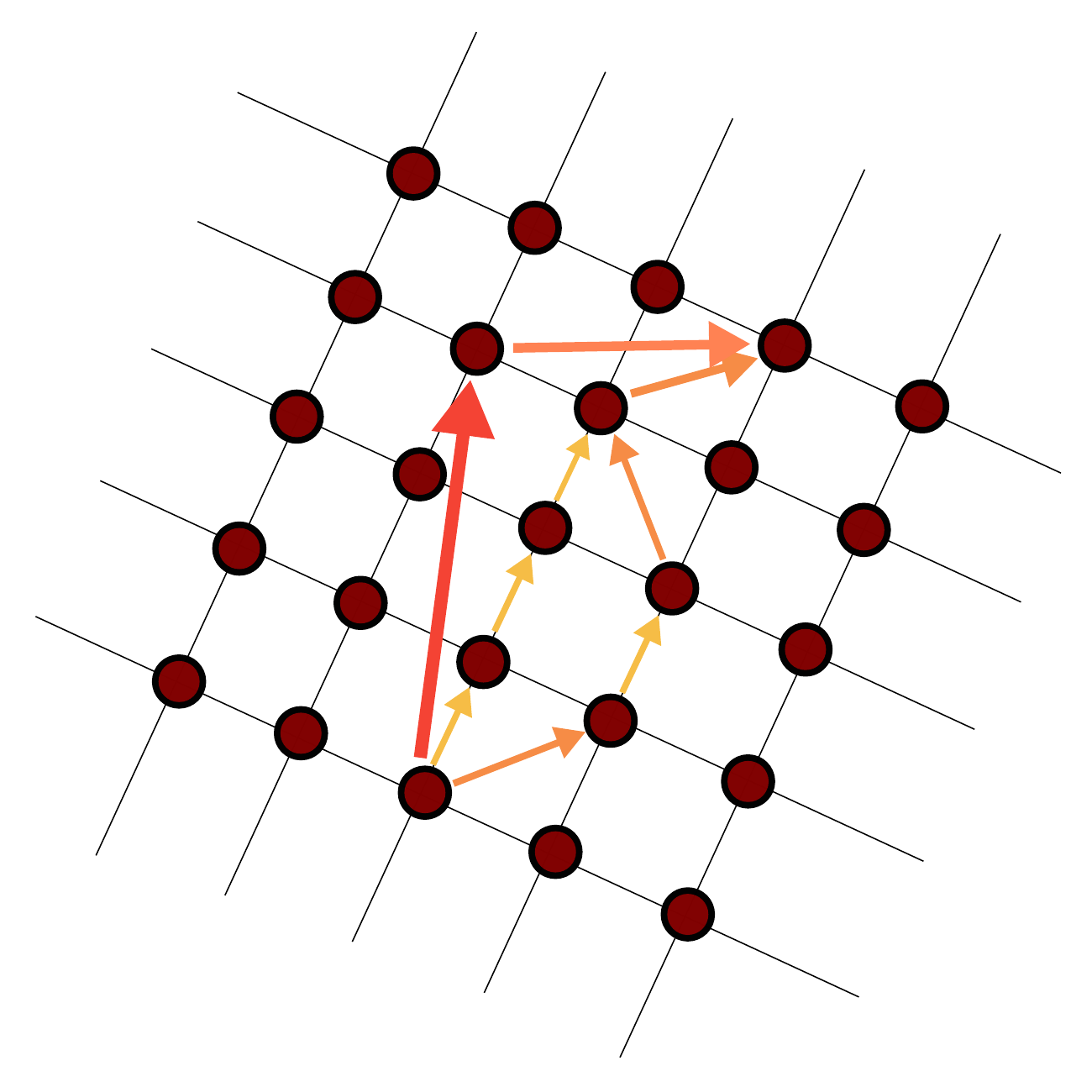}
    \caption{A sketch of multiple possible trajectories along domain wall transitions in a discrete landscape of isolated vacua. The redness of each arrow indicates the increasing charge carried by the corresponding domain wall.}
    \label{fig:DW_lattice}
\end{figure}

At this point, one may be tempted to simply define the distance between the vacua $|\mathbf{n}_1\rangle$ and $|\mathbf{n}_2\rangle$ as
\begin{eqaed}\label{eq:pre-definition}
    \Delta_\text{ss}(\mathbf{n}_1, \mathbf{n}_2) \equiv \int dr \, \sqrt{G_{ij}(\phi_\text{DW}(r)) \, \frac{d\phi^i_\text{DW}}{dr} \, \frac{d\phi^j_\text{DW}}{dr}} \, ,
\end{eqaed}
where the subscript stands for ``single step'', but this expression has a few drawbacks. To begin with, there may exist multiple domain walls interpolating between the same vacua. Borrowing terminology from flux compactifications for concreteness, $\mathbf{n}$ represents quantized charges or fluxes\footnote{Strictly speaking, in this scenario the reference EFT with this set of vacua is higher-dimensional, since the fluxes threading internal cycles are found solving higher-dimensional field equations, as opposed to minimizing a $d$-dimensional scalar potential.}, and a domain wall may carry all of the net charge in a single transition or split it into several transitions. It would therefore seem ambiguous whether one should compute the distance with a single transition or summing smaller steps. In the case of exact moduli spaces, this ambiguity in the trajectory is fixed naturally by choosing geodesics. Therefore it seems reasonable to keep our definition in the discrete as close as possible, at least in spirit, to the continuous case. The second drawback actually points to a resolution: in top-down examples of flux compactifications, the Planck scale of the EFT changes as fluxes are varied because the (stabilized) volume of the internal space depends on $\mathbf{n}$. Since the expression in~\cref{eq:SDC} contains the Planck scale, it seems that the most reasonable course of action to address both issues, at least partially, is to define the distance summing along elementary domain walls, which modify the charges by a basis vector of the lattice $\mathbf{n} \to \mathbf{n} + \mathbf{e}$. If there are multiple such paths, one can minimize over them. As result, we arrive at the definition
\begin{eqaed}\label{eq:DW_definition}
    \frac{\Delta(\mathbf{n}_1, \mathbf{n}_2)}{M_\text{pl}(\mathbf{n}_2)^{\frac{d-2}{2}}} \equiv \min_{\sum_j \mathbf{e}_j = \mathbf{n}_2 - \mathbf{n}_1} \sum_k \frac{\Delta_\text{ss}(\mathbf{n}_1 + \sum_{l < k} \mathbf{e}_l, \mathbf{n}_1 + \sum_{l \leq k} \mathbf{e}_l)}{M_\text{pl}(\mathbf{n}_1+\sum_{l\leq k} \mathbf{e}_l)^{\frac{d-2}{2}}}  \, ,
\end{eqaed}
which is meant to gradually ``adjust'' the Planck scale as the path moves in field space. The sum is over all steps in the chain of domain wall transitions, with the $k$-th step adding $\mathbf{e}_k$ to the preceding ``flux'', as depicted in~\cref{fig:DW_lattice}. This point of view was emphasized in~\cite{Basile:2022sda} attempting to define a suitable distance beyond exact moduli spaces.

While in spirit~\cref{eq:DW_definition} is meant to adhere as closely as possible to the standard framework of geodesics distances in exact moduli spaces, at first glance computing such an expression in concrete examples seems daunting. However, with the aid of some inequalities and estimates, one can obtain rather general bounds that avoid explicit computations. This is the focus of the next section.

\section{Bounds on BPS domain walls in supergravity}\label{sec:DW_bounds}

As we discussed, the general expression in~\cref{eq:DW_definition} is quite difficult to compute in practice, and requires a numerical approach. Even if the field profiles of any domain wall solution were known in closed form, the minimization procedure would require some form of automation in general. However, we are ultimately interested in infinite-distance limits, specifically testing whether the exponential falloff of tower mass scales persists when moduli spaces are lifted to isolated vacua. Hence, it may prove useful to avoid explicit computations and provide bounds for~\cref{eq:DW_definition} instead. A convenient setting to attempt such task is minimal supergravity in four dimensions, where BPS domain walls are stable and characterized by simpler differential equations of the \emph{first order}. In the notation of the preceding section, the BPS equations read~\cite{Cvetic:1992bf, Cvetic:1992st, Cvetic:1992sf, Cvetic:1993xe, Cvetic:1996vr, Ceresole:2006iq, Bandos:2018gjp, Bandos:2019wgy}
\begin{eqaed}\label{eq:BPS_DW_eqs}
    \frac{dA}{dr} & = \abs{Z} \, , \\
   \frac{d\phi^i}{dr} & = - \, 2 \, K^{i\bar{j}} \, \partial_{\bar{j}} \abs{Z} \, ,
\end{eqaed}
where the ``central charge'' $Z \equiv e^{\frac{K}{2}} W$ is defined by the K\"{a}hler potential $K$ and the superpotential $W$, and attains the values $Z_1 \, , \, Z_2$ at the two vacua. The cosmological constant of the resulting supersymmetric anti-de Sitter (AdS) vacua is $\Lambda = -3 \abs{Z}^2$. Up to irrelevant proportionality factors, which we shall not consider, the ``single step'' distance then simplifies to
\begin{eqaed}\label{eq:DW_ss_sugra}
    \Delta_\text{ss}(\mathbf{n}_1, \mathbf{n}_2) = \int dr \, \abs{\partial \abs{Z}} \, ,
\end{eqaed}
and involves the modulus of the gradient in field space defined by the metric $G$ induced by $K$. Here and in the following, symbols such as $\abs{\partial \abs{Z}}$ involve derivatives with respect to $\phi^i$. The expression in \cref{eq:DW_ss_sugra} springs an intriguing observation: if the square root in the integrand were \emph{removed}, the result would simply be the tension $T_\text{DW}$ of the domain wall, since $\abs{\partial \abs{Z}}^2$ is proportional to $\frac{d\abs{Z}}{dr}$~\cite{Ceresole:2006iq} (we take $\abs{Z}$ to be monotonically increasing for definiteness). On the other hand, as we will discuss in~\cref{sec:holography}, a natural holographic counterpart of the domain wall distance is the length of its dual RG flow\footnote{In non-supersymmetric settings with a metastable AdS configuration, the existence of different RG flows dual to vacuum bubbles has been proposed~\cite{Antonelli:2018qwz, Ghosh:2021lua}.}~\cite{Girardello:1998pd, Girardello:1999bd, Freedman:1999gp, Myers:2010xs, Myers:2010tj}. The analog of the quantum information distance~\cite{oconnor, Stout:2021ubb, Stout:2022phm} (see also~\cite{Anselmi:2011bp} for a related notion) with the square root removed was investigated in~\cite{Anselmi:2002fk}, where it was shown that the corresponding ``length'' yields the difference in $a$-central charges of the conformal field theories (CFTs) at the endpoints of the flow. This matches nicely with the known holographic dictionary, since $\abs{Z}$ in supersymmetric AdS vacua is in fact dual to the central charge~\cite{Ceresole:2006iq}. While this perspective is simple, independent of the specific field profiles $\phi_\text{DW}(r)$ and avoids essentially all of the computational difficulties of~\cref{eq:DW_definition}, we stick to the approach that, in our view, is closest in spirit to the established geometry of exact moduli spaces.

In order to find useful bounds from BPS domain walls, let us focus on the ``single step'' distance $\Delta_\text{ss}$ first. In the following, all arbitrary functions of $r$ are chosen to be sufficiently ``nice'' in the sense that all integrals converge.

\subsection{Upper bounds}\label{sec:upper_bounds}

Let us consider the Cauchy-Schwarz inequality
\begin{eqaed}\label{eq:cauchy-schwarz}
    \norm{f g}_1 \leq \norm{f}_2 \, \norm{g}_2 \, ,
\end{eqaed}
applied to the functions
\begin{eqaed}\label{eq:f_g_CS}
    f = \sqrt{\varphi} \, , \qquad g = \frac{\abs{\partial \abs{Z}}}{\sqrt{\varphi}}
\end{eqaed}
for some positive function of the form $\varphi(\abs{Z(r)})$. Therefore, up to the prefactor in~\cref{eq:DW_ss_sugra}, the left-hand side of~\cref{eq:f_g_CS} is $\Delta_\text{ss}$, and one obtains
\begin{eqaed}\label{eq:DW_ss_inequality_1}
    \Delta_\text{ss}^2 \leq \left(\int \varphi \, dr \right) \left( \int \abs{\partial \abs{Z}}^2 \, \varphi^{-1} \, dr \right) \, .
\end{eqaed}
This expression is useful, because the last integral can be rewritten (again up to an irrelevant prefactor) as
\begin{eqaed}\label{eq:phi_Z_trick}
    \int \abs{\partial \abs{Z}}^2 \, \varphi(\abs{Z})^{-1} \, dr \propto \int_{\abs{Z_1}}^{\abs{Z_2}} \varphi(\abs{Z})^{-1} \, d\abs{Z} \, ,
\end{eqaed}
which is independent of the field profile $\phi_\text{DW}(r)$ induced by the domain wall. Unfortunately we did not find a simple inequality that only involves such integrals, but the other one can be estimated as we shall now argue. Firstly, $\varphi$ must be chosen so that both integrals (which involve two different integration variables) converge. Linearizing the BPS equations around the AdS vacua, one finds that the approach to the fixed points $Z_1 \, , \, Z_2$ is exponential in $r$. Thus, one may pick
\begin{eqaed}\label{eq:phi_choice}
    \varphi(\abs{Z}) = \frac{1}{1 + \log^2\left(\frac{\abs{Z} - \abs{Z_1}}{\abs{Z_2} - \abs{Z}}\right)} \, ,
\end{eqaed}
which decays as $r^{-2}$ as a function of $r$ and whose reciprocal $\varphi^{-1}$ has integrable singularities as a function of $\abs{Z}$. As a result, the last integral in~\cref{eq:DW_ss_inequality_1} gives
\begin{eqaed}\label{eq:last_integral_inequality_1}
    \int_{\abs{Z_1}}^{\abs{Z_2}} \varphi^{-1}(\abs{Z}) \, d\abs{Z} = \left(1 + \frac{\pi^2}{3}\right) \Delta \abs{Z} \propto T_\text{DW} \, .
\end{eqaed}
All in all, we arrive at
\begin{eqaed}\label{eq:DW_ss_inequality_1_final}
    \Delta_\text{ss} \leq k \, \sqrt{T_\text{DW}} \left( \int \frac{dr}{1 + \log^2\left(\frac{\abs{Z} - \abs{Z_1}}{\abs{Z_2} - \abs{Z}}\right)} \right)^{\frac{1}{2}} \, ,
\end{eqaed}
where $k$ subsumes all numerical (and calculable) prefactors. In this derivation we have assumed that $\abs{Z}$ is regular, but in the case of flux vacua membranes can carry the missing charge along a transition and induce a discontinuity~\cite{Bandos:2018gjp}. In this case $\frac{d\abs{Z}}{dr} = 4 \, \abs{\partial \abs{Z}}^2 + \frac{1}{2} \, T_\text{M}$, where $T_\text{M} \leq T_\text{DW}$ is the membrane tension. As a result~\cref{eq:phi_Z_trick} has an additional \emph{negative} term proportional to $T_\text{M}$, which can be removed at the price of replacing the equal sign with $\leq$. Therefore, the final bound in~\cref{eq:DW_ss_inequality_1_final} still holds.

\subsubsection{Upper bound estimates}\label{sec:upper_estimates}

So far, the inequalities we derived hold up to numerical constants which do not depend on any parameters. These factors may be easily reinstated in the above expressions, but they are irrelevant since we are ultimately interested in large-distance scalings. In order to massage the above expressions into more useful bounds, we now estimate the integral in~\cref{eq:DW_ss_inequality_1_final} using the rough behavior of $\abs{Z(r)}$. Since the approach to the fixed points is exponentially fast,
\begin{eqaed}\label{eq:Z_exp_asymptotics}
    \abs{Z} & \overset{r \,\to\, +\infty}{\sim} \abs{Z}_+ \equiv \abs{Z_2} - A_2 \, e^{-\alpha_2 r} \, , \\
    \abs{Z} & \overset{r \,\to\, -\infty}{\sim} \abs{Z}_- \equiv \abs{Z_1} + A_1 \, e^{\alpha_1 r} \, , \\
\end{eqaed}

we approximate $\abs{Z} \approx \frac{\abs{Z_1} + \abs{Z_2}}{2}$ in a central region $r_- < r < r_+$ and $\abs{Z} \approx \abs{Z_\pm}$ outside, where $r_+ \equiv \max(0 \, , - \frac{1}{\alpha_2} \log \frac{\Delta \abs{Z}}{A_2})$ and $r_- \equiv \min(0 \, , \frac{1}{\alpha_1} \log \frac{\Delta \abs{Z}}{A_1})$ are chosen to make $\varphi(\abs{Z_\pm})$ well-defined. Hence, the approximated integral with the choice of~\cref{eq:phi_choice} is
\begin{eqaed}\label{eq:phi_integral_approx}
    \int \varphi(\abs{Z}) \, dr & = \int_0^{+\infty} \varphi(\abs{Z}) \, dr + \int_{-\infty}^0 \varphi(\abs{Z}) \, dr \\
    & \approx \int_{r_+}^{+\infty} \varphi(\abs{Z}_+) \, dr + \int_{r_-}^{r_+} 1 \, dr + \int_{-\infty}^{r_-} \varphi(\abs{Z}_-) \, dr \, .
\end{eqaed}
After a change of variables, the first and last terms simplify to ($i = 1 \, , \, 2)$
\begin{eqaed}\label{eq:asymp_integrals_simp}
    \frac{1}{\alpha_i}\int_0^{\min(1 \, , \, x_i)} \frac{1}{1 + \log^2 \left( \frac{u}{x_i - u} \right)} \frac{du}{u} \leq \frac{\pi}{2 \alpha_i}\, , \qquad x_i \equiv \frac{\Delta \abs{Z}}{A_i} \, .
\end{eqaed}
The remainder $r_+ - r_-$ can be upper-bounded by $\frac{1}{\alpha_2} \abs{\log \frac{\Delta \abs{Z}}{A_2}} + \frac{1}{\alpha_1} \abs{\log \frac{\Delta \abs{Z}}{A_1}}$. All in all, a weaker and approximate version of the upper bound of~\cref{eq:DW_ss_inequality_1_final} takes the simple form
\begin{eqaed}\label{eq:DW_ss_inequality_1_final_approx}
    \Delta_\text{ss} \lesssim \sqrt{\frac{T_\text{DW}}{\alpha_1} \left(\frac{\pi}{2} + \abs{\log \frac{\Delta \abs{Z}}{A_1}} \right) + \frac{T_\text{DW}}{\alpha_2} \left(\frac{\pi}{2} + \abs{\log \frac{\Delta \abs{Z}}{A_2}} \right)} \, .
\end{eqaed}

\subsubsection{Scalings at large fluxes}\label{sec:upper_bound_large_flux}

Let us now further assume that the stabilization arises from fluxes, as usual in compactifications arising from string theory or string-inspired models. At large flux, say along a specific direction $\mathbf{n} = n \, \mathbf{e}$ with $n \gg 1$, at AdS vacua one typically has $\abs{Z} \propto n^{-\gamma}$ for some $\gamma < 0$.

In order to estimate the scaling with $n$ of the approximate upper bound of~\cref{eq:DW_ss_inequality_1_final_approx}, we observe that for an elementary transition $\Delta \abs{Z} \sim \frac{d\abs{Z}}{dn} \propto n^{-\gamma - 1}$, while $\alpha_1 \sim \alpha_2$ arise linearizing the BPS equations for $\phi^i_\text{DW} \sim \phi_0^i + \delta \phi^i$ around the AdS vacua, which brings along the Hessian matrix of $\abs{Z}$ from~\cref{eq:BPS_DW_eqs}. Since $\frac{d}{dr}\abs{Z} \propto \abs{\partial \abs{Z}}^2 \propto \abs{\frac{d}{dr}\delta \phi}^2$, the exponents $\alpha_i$ are twice the maximum or minimum eigenvalues of the Hessian matrix of $\abs{Z}$, depending on whether $r \to +\infty$ or $r \to -\infty$.

If $\phi^i$ represent (pseudo-)moduli for volumes of internal (holomorphic) cycles in an ``electric'' frame, one typically expects them to scale as the Kaluza-Klein (KK) scale $\ell_\text{KK}^2$. For instance, the K\"{a}hler moduli $t^a$ of a Calabi-Yau compactification from ten dimensions to four dimensions satisfy $D_{abc} t^a t^b t^c = \text{Vol}(\text{CY}_3) \propto \ell_\text{KK}^6$, assuming no significant deviations from isotropy. The dual ``magnetic'' moduli $\tau_a = D_{abc} t^b t^c$ scale differently. Thus, if $m_\text{KK} \sim n^{-\beta}$ is the curvature scale of the internal space, with $\beta > 0$, a schematic estimate for the exponents $\alpha$ is $\abs{Z}/\phi_0^2 \propto n^{4\beta-\gamma}$. This leads to
\begin{eqaed}\label{eq:DW_ss_inequality_1_fluxes}
  \Delta_\text{ss} \lesssim n^{2\beta - \frac{1}{2}} \, \sqrt{\log n} \, .
\end{eqaed}
To connect to the formula of~\cref{eq:DW_definition}, we also need the relation $M_\text{pl}^{d-2}(\mathbf{n}) \propto \ell_\text{KK}^p$, where $p$ denotes the internal dimensions. Since this relation holds at the level of the $(d+p)$-dimensional EFT, we expect the supersymmetric case with $d=4$ to have $p \leq 6$ and even. According to~\cref{eq:DW_definition}, one needs to multiply by $\frac{1}{n^{\frac{\beta p}{2}}}$ and sum over $n$ from $n_1$ to $n_2$. The final result is the approximate upper bound
\begin{eqaed}\label{eq:DW_upper_flux_scaling}
    \frac{\Delta(n_1, n_2)}{M_\text{pl}(n_2)} & \lesssim \sum_{n=n_1}^{n_2} n^{(2- \frac{p}{2})\beta - \frac{1}{2}} \sqrt{\log n} \\
    & \underset{\beta \neq \frac{1}{p-4}}{\overset{n_2 \gg n_1}{\sim}} n_2^{(2 - \frac{p}{2})\beta + \frac{1}{2}} \sqrt{\log n_2} \, ,
\end{eqaed}
where the asymptotics is given up to a numerical prefactor. If $\beta = \frac{1}{p-4}$, the leading asymptotics is instead given by $\log^\frac{3}{2}n_2$. For definiteness, the interesting case $p=6$ gives $n_2^{-\beta+\frac{1}{2}} \sqrt{\log n_2}$ for $\beta \neq \frac{1}{2}$.

The scaling corresponding to a discrete version of the distance conjecture is logarithmic. To wit, since all masses are expected to scale with a power of $n$, the result would be an exponential decay in $\Delta/M_\text{pl}$. From~\cref{eq:DW_upper_flux_scaling} we see that there is a window for such scaling, insofar as the exponent of $n_2$ is non-negative. This means that, within the assumptions that led to this estimate, assuming the distance conjecture for ordinary moduli spaces holds, a discrete distance conjecture for BPS domain walls can only hold if
\begin{eqaed}\label{eq:beta_bound}
    (p-4)\beta < 1 \, ,
\end{eqaed}
which in the most interesting case $p=6$ of ten-dimensional $\mathcal{N}=1$ compactification entails $\beta < \frac{1}{2}$. As an example, the DGKT vacuum has $\beta = \frac{1}{4}$~\cite{DeWolfe:2005uu}, but the string coupling $g_s \propto n^{-\frac{3}{4}}$ also depends on $n$. Adjusting for this, the new bound on the discrete distance is $\Delta/M_\text{pl} \lesssim \sqrt{\frac{\log n_2}{n_2}}$, which would seem in tension with an exponential decay or even with the existence of an infinite-distance limit. However, before drawing conclusions from this result, let us emphasize that the DGKT construction contains other ingredients that should be taken into account, such as complex structure moduli (in the case of Calabi-Yau orientifolds) blow-up moduli (in the case of $T^6/\mathbb{Z}_3^2)$, smearing and/or additional fields, and these additional elements may modify the results. For instance, the inclusion of open-string (pseudo-)moduli in the EFT arising from D4-branes~\cite{Shiu:2022oti} leads to geodesic distances that do scale logarithmically with $n$. Indeed, these moduli do not have the same geometrical interpretation, and the above estimates cannot be applied anymore. After the field redefinition that brings the field space metric of~\cite{Shiu:2022oti} to the canonical one on $\mathbb{H}^2 \times \mathbb{R}$, the corresponding field scales has scaling exponent $\beta = 1$, which modifies the upper bound for the ``single step'' distance $\Delta_\text{ss}$ to $\Delta_\text{ss}/M_\text{pl} \lesssim \sqrt{\log n}$. Hence, the total distance $\Delta/M_\text{pl} \lesssim n_2 \, \sqrt{\log n_2}$, which is now compatible with the results of~\cite{Shiu:2022oti}.

Another, more general, consideration is that by AdS/CFT holography one expects the central charge of the dual CFT to scale as $c \propto \ell_\text{AdS}^2 \propto n^{2\gamma}$. If scale separation is absent\footnote{For some studies in this direction, see~\cite{Collins:2022nux, Lust:2022lfc}.} $\gamma = \beta$ and the EFT would be replaced by a consistent truncation. As a result, the bound $\beta < \frac{1}{2}$ would entail that the central charge $c \ll n$. Generally speaking, it seems difficult to realize a holographic central charge with this property. The typical behaviors of central charges can only be compatible with our analysis if $\gamma > \beta$, that is $\ell_\text{AdS} \gg \ell_\text{KK}$.

\subsection{Lower bounds}\label{sec:lower_bounds}

Let us now discuss lower bounds on the discrete distance, starting from the ``single step'' quantity $\Delta_\text{ss}$. This time we consider the reverse H\"{o}lder inequality,
\begin{eqaed}\label{eq:reverse-holder}
    \norm{f \, g}_1 \geq \norm{f}_{\frac{1}{q}} \, \norm{g}_{- \frac{1}{q-1}} \, , \qquad q > 1
\end{eqaed}
applied to the functions
\begin{eqaed}\label{eq:f_g_RH}
    f = \abs{\partial \abs{Z}}^{q+1} \, \varphi \, , \qquad g = \abs{\partial \abs{Z}}^{1-q} \, .
\end{eqaed}
Writing out the integrals explicitly, one obtains
\begin{eqaed}\label{eq:RHI_integrals}
    \int \abs{\partial \abs{Z}}^2 \, \varphi \, dr \geq \left( \int \abs{\partial \abs{Z}}^{1 + \frac{1}{q}} \, \varphi^{\frac{1}{q}} \, dr \right)^q \left( \int \abs{\partial \abs{Z}} \, dr \right)^{1-q} \, ,
\end{eqaed}
which gives the (family of) lower bound(s)
\begin{eqaed}\label{eq:DW_ss_lower_bound_fam}
    \Delta_\text{ss} \geq \left(\frac{\int \abs{\partial \abs{Z}}^{1 + \frac{1}{q}} \, \varphi^{\frac{1}{q}} \, dr}{\int \abs{\partial \abs{Z}}^2 \, \varphi \, dr}\right)^{\frac{1}{q-1}} \int \abs{\partial \abs{Z}}^{1 + \frac{1}{q}} \, \varphi^{\frac{1}{q}} \, dr \, .
\end{eqaed}
Taking $q \to +\infty$, a straightforward expansion gives the simpler bound
\begin{eqaed}\label{eq:DW_ss_lower_bound}
    \Delta_\text{ss} \geq e^{- \frac{\int \abs{\partial \abs{Z}}^2 \, \varphi \, \log(\abs{\partial \abs{Z}} \, \varphi) \, dr}{\int \abs{\partial \abs{Z}}^2 \, \varphi \, dr}} \int \abs{\partial \abs{Z}}^2 \, \varphi \, dr \, ,
\end{eqaed}
where once again the last factor rearranges into an integral that does not depend on the profile $\phi_\text{DW}(r)$ choosing $\varphi = \varphi(\abs{Z})$. The exponent can be interpreted as the average $\langle \log(\abs{\partial \abs{Z}} \, \varphi)\rangle_\rho$ with respect to the probability distribution with (non-normalized) density $\rho = \abs{\partial \abs{Z}}^2 \, \varphi$, and thus letting $M \equiv \sup_r \abs{\partial \abs{Z}} \, \varphi$, if $M$ is finite\footnote{If $\abs{Z}$ is regular, all useful choices of $\varphi$ give a finite $M$, since $\abs{\partial \abs{Z}}$ decays exponentially at infinity. In the case of fluxes, the presence of membranes carrying charges can induce discontinuities. A way to treat them is to introduce a ``flowing superpotential that incorporates the jump in charge(s)\emph{e.g.}~\cite{Bandos:2018gjp}.} one has
\begin{eqaed}\label{eq:avg_trick}
    e^{- \langle \log(\abs{\partial \abs{Z}} \, \varphi) \rangle_\rho} \geq e^{- \log M} = \frac{1}{M} \, .
\end{eqaed}
Therefore, up to a numerical (and calculable) prefactor $k$, one arrives at the weaker bound
\begin{eqaed}\label{eq:DW_ss_simpler_lower_bound}
  \Delta_\text{ss} \geq \frac{k}{M} \int_{\abs{Z_1}}^{\abs{Z_2}} \varphi(\abs{Z}) \, d\abs{Z} \, .  
\end{eqaed}
Once again, this bound holds assuming $\abs{Z}$ is regular. In the presence of discontinuities induced by membranes, these steps are not well-defined due to the appearance of $\delta$-distributions in rational powers or logarithms.

\subsubsection{Lower bound estimates and flux scalings}\label{sec:lower_estimates}

In order to estimate $M$ precisely one would need to know either the width or the peak slope of the kink profile of $\abs{Z}$. One expects $\abs{\partial \abs{Z}}$ to be sharply peaked and decay exponentially, and thus choosing a slowly-varying (or outright constant) $\varphi$ would allow estimating $M$ with the height of the peak. By contrast, choosing $\varphi$ to be a window function that vanishes around the peak would avoid the need to know its height, but this requires knowing its width.

To obtain some qualitative insights from~\cref{eq:DW_ss_simpler_lower_bound}, we take the first approach and estimate the height of the peak in $\abs{\partial \abs{Z}}$ as the average slope $\frac{\abs{\Delta \abs{Z}}}{\abs{\Delta \phi}} \propto \frac{T_\text{DW}}{\abs{\Delta \phi}}$. Choosing a constant $\varphi$ then simplifies~\cref{eq:DW_ss_simpler_lower_bound} to $\Delta_\text{ss} \gtrsim \abs{\Delta \phi}$.

In terms of the flux scalings discussed in~\cref{sec:upper_bound_large_flux}, $\abs{\Delta \phi} \sim \abs{\frac{d\phi}{dn}} \sim n^{2\beta - 1}$. Putting together the lower bound of~\cref{eq:DW_ss_simpler_lower_bound} and the upper bound of~\cref{eq:DW_ss_inequality_1_fluxes}, one arrives at the approximate inequalities
\begin{eqaed}\label{eq:double_bound}
    n^{2\beta-1} \lesssim \Delta_\text{ss} \lesssim n^{2\beta - \frac{1}{2}} \, \sqrt{\log \, n} \, ,
\end{eqaed}
which can then be turned into a bound for the bulk total distance according to~\cref{eq:DW_definition} as in~\cref{eq:DW_upper_flux_scaling}. Dividing by the Planck mass changes the exponents in the lower bound as in~\cref{eq:DW_upper_flux_scaling}, and since $\beta > 0$ the resulting expression $n^{(2-\frac{p}{2})\beta - 1}$ would suggest that, in order for a discrete distance conjecture to hold, the number of extra dimensions has to be at $p \geq 4$. Otherwise, the scaling of the lower bound would exclude a logarithmic behavior $\Delta_\text{ss}/M_\text{pl} \sim n^{-1}$, or equivalently $\Delta(n_1, n_2) \sim \log \frac{n_2}{n_1}$. Of course the validity of these considerations is somewhat shaky, since the estimates beyond the sharp bounds derived from H\"{o}lder's inequalities are not under control quantitatively.

\section{Examples on hyperbolic field spaces}\label{sec:examples}

As we have seen, the bounds that we were able to derive on general grounds are not very constraining, including the simpler estimates based on large-flux scalings. It is instructive to consider more concrete examples, in which one can derive more stringent bounds.

We will focus on hyperbolic field spaces, which are ubiquitous in supersymmetric compactifications due to the properties of nilpotent orbits~\cite{Grimm:2018ohb}. The fields arrange into pairs of axions and saxions $(a,s)$ with line element
\begin{eqaed}\label{eq:hyperbolic_line_element}
    ds^2_{\mathbb{H}^2} = \frac{da^2 + ds^2}{s^2} \, .
\end{eqaed}
The ``single step'' distance $\Delta_\text{ss}$ can be now bounded in a much simpler manner, using the inequalities $\max(\abs{x} \, , \, \abs{y}) \leq \sqrt{x^2 + y^2} \leq \abs{x} + \abs{y}$. Taking the field profiles to be increasing for definiteness, one obtains
\begin{eqaed}\label{eq:DW_ss_hyperbolic_bound}
    \max \left(\int \frac{da}{s} \, , \, \log \frac{s(\mathbf{n_2})}{s(\mathbf{n_1})} \right) \leq \Delta_\text{ss} \leq \int \frac{da}{s} + \log \frac{s(\mathbf{n_2})}{s(\mathbf{n_1})} \, .
\end{eqaed}
This bound can be weakened and simplified by taking maximum and minimum values of $s$ in the integrals, which yields
\begin{eqaed}\label{eq:DW_ss_hyperbolic_bound_weaker}
    \max \left( \frac{\Delta a}{s(\mathbf{n_2})} \, , \, \log \frac{s(\mathbf{n_2})}{s(\mathbf{n_1})} \right) \leq  \Delta_\text{ss}(\mathbf{n_1}, \mathbf{n_2}) \leq \frac{\Delta a}{s(\mathbf{n_1})} + \log \frac{s(\mathbf{n_2})}{s(\mathbf{n_1})} \, .
\end{eqaed}
For ordinary trajectories dominated by saxions, at large moduli space distances $s \to +\infty$ as a power $n^\alpha$ while $\Delta a \ll 1$. Thus, if $\alpha \geq 1$ the logarithmic terms dominate, since
\begin{eqaed}\label{eq:log_saxion_ss}
    \log \frac{s((n+1) \, \mathbf{e})}{s(n \, \mathbf{e})} \sim \frac{\alpha}{n} \, ,
\end{eqaed}
and the domain wall distance scales as the geodesic distance. Instead, if $\alpha < 1$ the axionic terms can dominate if $\Delta a = \mathcal{O}(1)$, and the discrete distance conjecture would not follow from the ordinary one. This is actually an known result in a new guise: this kind of non-geodesicity for hyperbolic field spaces was explored in~\cite{Calderon-Infante:2020dhm}\footnote{Similar settings were considered in~\cite{Freigang:2023ogu} in a cosmological context.} and was referred to a the \emph{swampy case}.

\begin{figure}[ht!]
    \centering
    \includegraphics[scale=0.5]{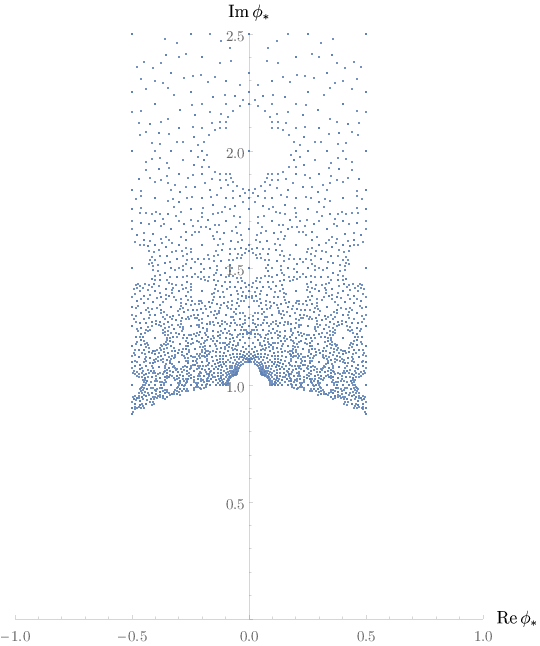}
    \caption{The discrete landscape of isolated vacua in the simple model above~\cref{eq:toy_W}. The plot is taken from the original reference~\cite{Bandos:2018gjp}.}
    \label{fig:toy_landscape}
\end{figure}

In the spirit of applying our considerations beyond string compactifications, let us also consider the toy model studied in~\cite{Bandos:2018gjp}. The model contains a single complex modulus $\phi$ with $\Im \phi > 0$, the bosonic component of a chiral superfield, with hyperbolic K\"{a}hler potential $K = - \, \log \Im \phi$, and a linear flux superpotential
\begin{eqaed}\label{eq:toy_W}
    W = (e_0 + i \, m^1) - i \, (e_1 + i \, m^0) \, \phi
\end{eqaed}
containing two electric integer fluxes $e_0 \, , \, e_1$ and two magnetic integer fluxes $m^0 \, , m^1$. The model has a discrete landscape of vacua (depicted in~\cref{fig:toy_landscape}), which is most conveniently parametrized by the complex fluxes $\alpha_0 \equiv e_0 - i \, m^1$ and $\alpha_1 \equiv e_1 - i \, m^0$. The field stabilizes at
\begin{eqaed}\label{eq:toy_stabilized}
    \phi_0 = i \, \frac{\alpha_0}{\alpha_1} \, ,
\end{eqaed}
with the constraint $\Im \phi_0 = \frac{\Re \alpha_0 \overline{\alpha_1}}{\abs{\alpha_1}^2} > 0$. It is possible to go to infinite distance in field space via elementary domain wall transitions if the fluxes in one of the two complex combinations $\alpha$ scales with $n$, which means that, up to $SL(2,\mathbb{Z})$ inversions, the exponent $\beta = 1$. This model contains no explicit reference to a flux-dependent Planck scale, and in this case the bounds of~\cref{eq:double_bound} would seem to exclude a logarithmic scaling, at least insofar as the estimates behind~\cref{eq:double_bound} are reliable. However, in this model the domain walls carrying the missing charge feature a discontinuity, which is induced by a membrane of tension $T_\text{M} \leq T_\text{DW}$. As a result, as discussed in~\cref{sec:lower_bounds}, the lower bound~\cref{eq:DW_ss_simpler_lower_bound} does not apply. To understand more intuitively what is happening, the lower-bound estimate in~\cref{eq:double_bound} relies on the finiteness of $M$ in~\cref{eq:DW_ss_simpler_lower_bound}, but since $\abs{Z}$ is discontinuous at the origin one can think of a ``divergent $M$'' trivializing the bound. Indeed, as shown in~\cite{Bandos:2018gjp}, there are elementary domain walls between vacua that are ``vertically aligned'', in the sense of~\cref{fig:toy_landscape}. Along such transitions only the saxions flow between flux minima according to $s_0(n) \to s_0(n) + 1$, implying that $\Delta_\text{ss} \sim \frac{1}{n}$ up to a factor. This recovers the geodesic logarithmic scaling of the full distance $\Delta(n_1, n_2) \sim \log \frac{n_2}{n_1}$. This should be the universal scaling at large distances in this model, because of the considerations on axion variations discussed above.

\subsection{Replacing hyperbolic space with flat space}\label{sec:flat_space_case}

In order to explore the relevance of hyperbolic field spaces in the above example, let us modify the model replacing it with a flat field space, with
\begin{eqaed}\label{eq:flat_line_element}
    ds^2_{\mathbb{H}^2} \quad \longrightarrow \quad da^2 + ds^2 \, .
\end{eqaed}
The corresponding K\"{a}lher potential is $K = \abs{\phi}^2$. Thus, supersymmetric vacua obey
\begin{eqaed}\label{eq:flat_SUSY_vacua}
    \partial W + \partial K \, W = -i \, \overline{\alpha_1} + \overline{\phi} \, W = 0 \, ,
\end{eqaed}
and for large fields (driven by some fluxes going to infinity) the first term is negligible. Hence, the solutions asymptote to $\phi \sim -i \, \frac{\overline{\alpha_0}}{\overline{\alpha_1}}$. To ensure that the field be self-consistently large, we take \emph{e.g.} $\alpha_1 = 1$ and $\alpha_0 = - i \, n$. We also need the first correction to compute the leading on-shell superpotential, and it reads
\begin{eqaed}\label{eq:flat_vev}
    \phi \sim -i \, \frac{\overline{\alpha_0}}{\overline{\alpha_1}} + i \, \frac{\alpha_1}{\alpha_0} = n - \frac{1}{n} \, .
\end{eqaed}
For this choice of fluxes, the other solution is not large for $n \gg 1$. The distance along increasing $n$ is now $\Delta \phi \sim n$, while the central charge
\begin{eqaed}\label{eq:flat_cc}
    \abs{Z} = e^K \abs{W} \sim \frac{1}{n} \, e^{n^2 - 2} \, .
\end{eqaed}
While for both flat and hyperbolic field spaces the cosmological constant grows in the large-$n$ limit, the exponential scaling with the distance is now replaced by $\exp \Delta^2$. This means that a modified model where the distance conjecture is satisfied would likely become inconsistent replacing the hyperbolic field space with a flat one, since it would bring along a factor $e^{\abs{\phi}^2}$ in the cosmological constant and dramatically change the distance scalings.

All in all, we see that EFTs with hyperbolic field spaces, at least insofar as the Planck scale does not strongly depend on $n$ (perhaps with stabilization induced by a potential), appear to be especially compatible with a discrete distance conjecture. It seems easy to violate the required scalings if the field spaces are not hyperbolic or the fields do not scale logarithmically with $n$. Typically, geometric moduli which appear in hyperbolic field spaces scale as powers of $n$ at stabilized points, while the dilaton is canonically normalized but typically stabilizes logarithmically in $n$. This phenomenon has been observed in non-supersymmetric compactifications as well~\cite{Basile:2020mpt, Basile:2022zee}, where the discrete distance still scales logarithmically due to bounds very similar to~\cref{eq:DW_ss_hyperbolic_bound_weaker}. If the discrete version of the distance conjecture will be put on a more solid footing, these hints may point to a universal structure of consistent (supersymmetric) EFTs which seems to nicely match the top-down understanding gathered in~\cite{Grimm:2018ohb, Corvilain:2018lgw, Grimm:2019wtx, Grimm:2019bey, Grimm:2019ixq, Grimm:2021ckh}.

\newpage

\section{Holographic RG flows}\label{sec:holography}

There is a natural holographic counterpart of the ideas that we have discussed. (BPS) domain walls between (supersymmetric) AdS vacua are typically associated with RG flows between the corresponding dual CFTs. The quantum information metric~\cite{Provost:1980nc, oconnor, Ruppeiner:1995, Ruppeiner:1996, Maity:2015rfa, Beny:2012qh, Stout:2022phm} is a natural extension of the Zamolodchikov metric~\cite{zamolodchikov} which allows, for example, to compute distances between theories along RG flows~\cite{Basile:2022zee, Basile:2022sda}. There is also a slightly different approach in the same spirit\footnote{Additional references where geometrical methods were employed to study RG flows are~\cite{Lassig:1989tc, Dolan:1993cf, Dolan:1995zq, Kar:2001qm, Kar:2002wx, Anselmi:2002fk}.}, described in~\cite{Anselmi:2011bp}, where the renormalization group scale is taken into account explicitly in the metric.

A holographic test of the discrete distance conjecture involves computing this distance summing over ``single step'' RG flows, dividing by the $C_T$ central charge defined by the stress-tensor correlator in order to account for bulk Planck units~\cite{Baume:2020dqd, Basile:2022sda}. If the resulting distance turns out to scale as $\log \, N$, with $N \gg 1$ the number of colors in the dual gauge theory (more generally some power of the central charge related to $n$ in the bulk), the discrete distance conjecture also reproduces the ``AdS distance conjecture'' of~\cite{Lust:2019zwm} using a notion of distance that takes into account the isolation of vacua.

\subsection{Test in \texorpdfstring{$\mathcal{N}=4$ super Yang-Mills}{N=4 super Yang-Mills}}\label{sec:N=4_SYM}

\begin{figure}[ht!]
    \centering
    \includegraphics[scale=0.5]{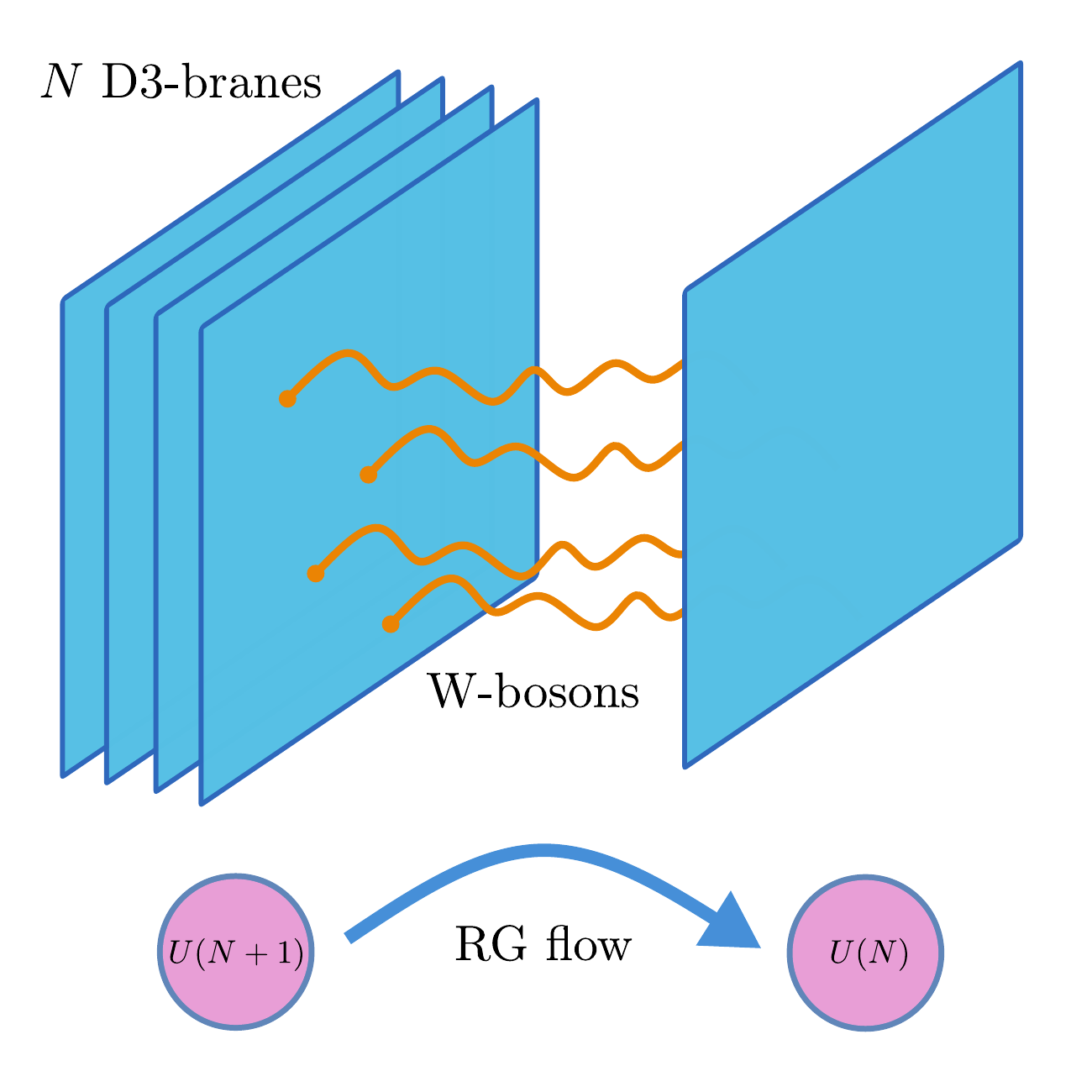}
    \caption{A sketch of the RG flow triggered by Higgsing the $U(N+1)$ gauge theory to $U(N)$, along with its bulk counterpart with separated D3-branes connected by W-bosons.}
    \label{fig:D3-brane_RG}
\end{figure}

The simplest setting to assess our proposal from a dual perspective is $U(N)$ $\mathcal{N}=4$ super Yang-Mills (SYM) and its dual bulk supergravity in AdS$_5$. The elementary domain walls interpolate between the vacuum with flux quanta $N$ and $N+1$. On the CFT side, this should correspond to an RG flow interpolating between $U(N)$ and $U(N+1)$. The microscopic picture consists of separating one out of $N+1$ coincident D3-branes from the remaining stack of $N$, Higgsing the gauge theory. The setup is depicted in~\cref{fig:D3-brane_RG}.

In~\cite{Gubser:1998iu, Gubser:1998kv, Intriligator:1999ai, Costa:2000gk}, the authors describe and conjecture various features of the putative deformation that implements the flow driven by Higgsing. Since the bare Lagrangian does not change, while the vacuum does, this deformation ought to arise integrating out the massive W-bosons corresponding to open strings stretching between the branes. In order to keep these modes in the theory, the correct decoupling limit is argued in~\cite{Costa:2000gk} to be slightly different than usual. The dual operator $\mathcal{O}_8$ is argued to be the (supersymmetric completion of the) first Born-Infeld correction to the Yang-Mills Lagrangian (see~\cite{Caetano:2020ofu} for more details),
\begin{eqaed}
   \mathcal{O}_8 = \mathcal{Q}^4 \overline{\mathcal{Q}}^4 \, \text{Tr} X^4 \propto \text{Str}\left(F^4 - \frac{1}{4} \, (F^2)^2\right) + \dots
\end{eqaed}
and it is argued that its dimension is \emph{exactly} 8. Therefore, the corresponding coupling $h$ has dimension \emph{exactly} -4 along the RG flow, which connects two superconformal fixed points breaking half of the supersymmetries (just like separating branes in the bulk does with respect to the near-horizon limit). The gauge-coupling beta function vanishes, and the anomalous dimension is therefore exact. The beta function of the $h$ coupling (defined by $\Delta\mathcal{L} = h \frac{\mathcal{O}_8}{\Lambda_4}$) is thus $\beta_h \sim 4h$, so that $h \sim h_0 \, e^{-4t}$ as a function of the RG time $t \equiv \log \frac{\Lambda}{\mu}$ in the infrared (IR) part of the flow $t \gg 1$, where the deformation operator is irrelevant and described by~\cref{eq:born-infeld_op}.

In order to evaluate the (asymptotic) distance along the RG flow, we need the two-point correlator of the deformation operators. In~\cite{Liu:1999kg} it was argued that the properly normalized correlator, namely the one associated to the usual Lagrangian $\mathcal{L}_\text{SYM} = -\frac{1}{4g_\text{YM}^2} \text{Tr} F^2 + \dots$, at the superconformal fixed point
\begin{eqaed}\label{eq:born-infeld_op}
    \langle \mathcal{O}_8(x) \mathcal{O}_8(0)\rangle \propto \frac{N^0 \, \lambda^4}{\abs{x}^{16}}
\end{eqaed}
does not scale with $N$ in the 't Hooft limit. The quantum information metric tensor is essentially a matrix of two-point correlators, for instance the prescription of~\cite{oconnor} defines
\begin{eqaed}\label{eq:oconnor_metric}
    g_{ab} = \int d^4x \, \langle \partial_a \mathcal{L}(x) \, \partial_b \mathcal{L}(0) \rangle
\end{eqaed}
for Lagrangians linear in the couplings $\lambda^a$. When pulled back along the RG flow, the metric depends on the energy-momentum trace $\Theta(x) = \beta_h \, \mathcal{O}(x)$, and one thus finds
\begin{eqaed}\label{eq:SYM_line_element}
   ds^2 \overset{\text{IR}}{\sim} e^{-8t} \, dt^2
\end{eqaed}
up to a prefactor of order $N^0$. The information distance is thus finite and of order $N^0$ across a single RG step $N \to N \pm 1$.

In order to build the correctly normalized bulk distance, one ought to divide by
\begin{eqaed}\label{eq:AdS5_planck_units}
    (L_{\text{AdS}} M_{\text{Pl}})^{\frac{3}{2}} = \sqrt{C_T(N)} \propto N \, .
\end{eqaed}
All in all, the distance across one step is proportional to $1/N$, and therefore the total bulk distance
\begin{eqaed}\label{eq:SYM_bulk_distance}
    \Delta(N_1, N_2) \sim \sum_{n=N_1}^{N_2} \frac{1}{\sqrt{C_T(n)}} \sim \log\frac{N_2}{N_1} \, .
\end{eqaed}
This is precisely the scaling predicted by the AdS distance conjecture, since the light Kaluza-Klein modes associated to the internal $S^5$ have masses that scale with powers of $N$ and thus decay exponentially in the properly normalized distance. Let us stress that this computation has the limitation of only including the IR contribution to the RG length of each step in the chain, where one has computational control. However, nothing seems to depend on $N$ prior to the normalization in~\cref{eq:SYM_bulk_distance}. Hence, integrating along the whole flow should not change the asymptotics which we obtain.

Another observation along the lines of~\cite{Basile:2022zee} is that $N=1$ does not seem to lie at infinite distance. This resonates with the idea that, since $U(1)$ SYM is a genuinely interacting CFT, it has no tower of conserved higher-spin currents. Had this not been the case, light higher-spin excitations would have had to replace the absent Kaluza-Klein modes in this regime~\cite{Lee:2018urn, Lee:2019wij, Lee:2019xtm}, but since there are none the distance cannot be finite on account of the emergent string proposal.

\subsubsection{Another large-\texorpdfstring{$N$}{N} limit}\label{sec:another_limit}

The above computation is carried out in the 't Hooft limit, where $N \gg 1$ with $\lambda \equiv g_{\text{YM}}^2 N = g_s N$ is kept fixed. However, one may object that we really should move along the values of $N$ without deforming any marginal couplings. Let us now analyze this case. Letting $k_8$ be the normalization of the correlator
\begin{eqaed}\label{eq:O8_correlator_another_limit}
    \langle \mathcal{O}_8(x) \mathcal{O}_8(0)\rangle = \frac{k_8}{\abs{x}^{16}}
\end{eqaed}
at the (IR) superconformal fixed point, the line element along the RG flow in the deep IR, where this expression is expected to be reliable, is proportional to $ds \overset{\text{IR}}{\sim} k_8 \, h_0 \, e^{-4t} \, dt$. In~\cite{Costa:1999sk, Costa:2000gk, Rastelli:2000xj} it is argued that $k_8 \propto \lambda^4$ which now scales as $N^4$ while the $\mathcal{O}(1)$ coefficient $h_0 \propto \frac{R_\text{sep}^4}{\delta^4}$ depends on the $\text{AdS}$ radius associated to the separated brane stack, something of order $N^0$ in our limit, and on the distance $\delta$ between the brane stacks.

In~\cite{Costa:1999sk} the authors describes a special scale, called \emph{gravitational gap}, given by $m_g \propto m_\text{W}/\sqrt{\lambda}$, where $m_W = g_\text{YM} \, \delta/\alpha'$ is the W-boson mass\footnote{This is the result in Einstein frame. In string frame there is no $g_\text{YM}$, but this difference is irrelevant in the limit we are now considering.}. Analyzing the holographic matching to the low-energy bulk EFT, the authors of~\cite{Costa:1999sk} conclude that $m_g$ is the physical UV cutoff of the EFT. If that is the case, the spacetime integral in~\cref{eq:oconnor_metric} is cut off at $m_g$ instead of $\Lambda$ (which may be taken to be $m_\text{W}$), bringing along a factor of $m_g^4/m_\text{W}^4 = \lambda^{-2}$ that cancels the dependence on $N$. As a result, the total bulk distance is again logarithmic. This effect is somewhat reminiscent of how the quantum gravity cutoff in the presence of parametrically many light species scales differently, but it is not clear whether a deeper connection with this result exists, especially since the standard dictionary uses the 't Hooft limit.

\subsection{Bulk perspective: separated D3-branes}

Let us now see what happens in the bulk dual of our $\mathcal{N}=4$ setup, namely the backreacted supergravity geometry produced by the D3-branes. This setting is particularly convenient, since the domain wall solutions we are interested in stem from ten-dimensional type IIB geometries that in the Einstein frame take the form
\begin{eqaed}\label{eq:10d_geometry}
   ds^2_{10} = H^{-1/2} \, dx_{3,1}^2 + H^{1/2} \, dy_6^2 \, ,
\end{eqaed}
where the transverse harmonic function (we omit the +1 in the near-horizon limit)
\begin{eqaed}\label{eq:harmonic_function}
   H(Y) = R^4 \, \int d^6y \, \frac{\sigma(y)}{\abs{Y-y}^4}
\end{eqaed}
describes the distribution of D3-branes, with $R^4 \propto \lambda$ the near-horizon AdS radius, so that $\int d^6y \, \sigma(y) = 1$. For general $\sigma$ the relation between the ten-dimensional solution and the 42 (gauged) supergravity scalars in five dimensions is very complicated, and appears to be only known for in a few special cases. However, for $SO(6)$-symmetric geometries only the radion $\rho$ is excited: writing
\begin{eqaed}\label{eq:transverse_metric}
   dy_6^2 = dr^2 + r^2 \, d\Omega^2_5 \, ,
\end{eqaed}
for $H = H(r)$ the only modulus that is excited is the overall volume of the sphere. Therefore, since also the dilaton does not vary along a domain wall transition, the only field that flows is the radion $\rho$, whose kinetic metric is $\frac{d\rho^2}{\rho^2}$ up to a prefactor.

Starting from $N_1$ D3-branes placed at the origin, $N_1 \, \sigma_\text{initial}(y) = N_1 \, \delta^{(6)}(y)$, separating $N_1 - N_2$ of them by an amount $\Delta$ and smearing them on the sphere gives
\begin{eqaed}\label{eq:smeared_branes}
   N_1 \, \sigma(y) = N_2 \, \delta^{(6)}(y) + \frac{N_1 - N_2}{\text{Vol}(S^5)} \, \int_{S^5} d\Omega(\hat{\nu}) \, \delta^{(6)}(y - \Delta \, \hat{\nu}) \, ,
\end{eqaed}
where the integral is over unit vectors $\hat{\nu}$. This distribution produces a radially symmetric harmonic function $H(r)$ with a 5d geometry that interpolates between $\text{AdS}_5(N_2)$ as $r \to 0$ and $\text{AdS}_5(N_1)$ as $r \to +\infty$.

With this setup, the distance across the DW is simply (up to a pre-factor)
\begin{eqaed}\label{eq:radion_distance}
   \int_{\rho(r=0)}^{\rho(r\to\infty)} \frac{d\rho}{\rho} = \log \frac{\rho(r\to\infty)}{\rho(r=0)} \propto \log \frac{H(r \to \infty)}{H(r \to 0)} = \log \frac{N_1}{N_2} \, .
\end{eqaed}
In this case, one can compute the total distance directly, since in the 't Hooft limit appropriate for holography the five-dimensional Planck mass does not depend on $N$.

The above computation seems to work for any $SO(6)$-symmetric distribution of branes. In particular, one can use a single domain wall carrying all of the charge or many elementary ones. This seems reasonable, since in this case there is only one flux and thus only one possible path. Perhaps in transitions where many scalars flow the specific distribution of branes will be more important, and there may be an optimal distribution that minimizes the total distance along the lines of~\cref{eq:DW_definition}.

\subsubsection{Separating without smearing}\label{sec:no_smearing}

As we have mentioned, extending the above computation beyond isotropic brane distributions is considerably more difficult, since it involves exciting more of the 42 scalar fields of the underlying five-dimensional gauged supergravity. However, for simple cases such as two separated brane stacks one can still bound the distance and show the relevant logarithmic scaling. This is because many distributions of D3-branes, encoded in~\cref{eq:harmonic_function}, only excite the diagonal components $\{ X_i \}_{i=1,\dots,6}$ of the unimodular $6 \times 6$ matrix of scalar fields appearing in the ten-dimensional metric~\cite{Cvetic:1999xx, Cvetic:2000nc, Taylor:2001fe, Bergshoeff:2004nq}. These fields describe deviations from sphericity, while the radion $\rho$ describes the change in volume. Therefore, on account of the discussion in the preceding section, in order to show that the total distance scales logarithmically one only need show that the contribution from the $X_i$ is bounded above by a logarithm. The kinetic terms take the form
\begin{eqaed}\label{eq:anisotropy_kinetic_terms}
    \sum_{i=1}^6 \frac{(\partial X_i)^2}{X_i^2} \, ,
\end{eqaed}
and thus the contribution to the (single-step) distance along the domain wall profile can be bounded above by
\begin{eqaed}\label{eq:anisotropy_upper_bound}
    \Delta_{\text{ss}}^{X} \leq k \, \sum_{i=1}^6 \log \frac{X_i^{\text{max}}}{X_i^{0}} \, ,
\end{eqaed}
where in the vacuum $X_i^{0} = 1$ by unimodularity and isotropy, up to an irrelevant $\mathcal{O}(1)$ constant $k$.

In order to see that the maximum value $X_i^{\text{max}}$ attained by the anisotropy fields scales at most as a power of $N$, we consider the simple case of two parallel brane stacks, without smearing. The corresponding harmonic function in~\cref{eq:harmonic_function} is given by
\begin{eqaed}\label{eq:unsmeared_branes}
   N_1 \, \sigma(y) = N_2 \, \delta^{(6)}(y) + (N_1 - N_2) \, \delta^{(6)}(y - \Delta \, \hat{\nu}) \, ,
\end{eqaed}
where without loss of generality we choose the direction $y \cdot \hat{\nu} = y_1$. The corresponding expression for $H$ depends separately on $y_1$ and $Y^2 \equiv y^2 - y_1^2$. As a result of this choice of preferred direction, $X_i = X_2$ for $i>1$, so that enforcing unimodularity the $X_i$ are given by
\begin{eqaed}\label{eq:unimodular_ansatz}
    X & = \left(X_1 \, , \, X_2 \, , \, X_2 \, , \, X_2 \, , \, X_2 \, , \, X_2 \right) \\
    & = (h_1 \, h_2^5)^{\frac{1}{6}} \left(\frac{1}{h_1} \, , \, \frac{1}{h_2} \, , \, \frac{1}{h_2} \, , \, \frac{1}{h_2} \, , \, \frac{1}{h_2} \, , \, \frac{1}{h_2} \right) \, ,
\end{eqaed}
following primarily the notation of~\cite{Bergshoeff:2004nq}. These fields appear in the harmonic function $H$ that defines the metric, to be matched to~\cref{eq:harmonic_function}. In order to match the two, one has to choose a coordinate system where the transverse space metric is the Euclidean $dy_6^2$ of~\cref{eq:10d_geometry}. This can be done as in~\cite{Bergshoeff:2004nq}, passing from the conventional coordinates $\mu^i$ of~\cite{Cvetic:1999xx, Cvetic:2000nc, Taylor:2001fe, Bergshoeff:2004nq} (subject to $\sum_i \mu^i \mu^i = 1$) to the Cartesian coordinates $y^i = \sqrt{h_i} \, \mu^i$ in
\begin{eqaed}\label{eq:H_bergshoeff}
    \frac{1}{H} = (h_1 h_2^5)^{\frac{1}{2}} \sum_i \frac{(\mu^i)^2}{h_i}
\end{eqaed}
The resulting expression for $H$ is a rational function of $y_1^2 \, , \, Y^2 \, , \, h_1 \, , \, h_2$, and the constraint
\begin{eqaed}\label{eq:y_constraint}
    \frac{y_1^2}{h_1} + \frac{Y^2}{h_2} = 1
\end{eqaed}
can be solved to eliminate $h_1$ or $h_2$ and simplify it. In the isotropic case $h_1 = h_2$ one recovers the familiar $H = (y_1^2 + Y^2)^{-2}$, in units of the radius $R$ of the $S^5$. In the anisotropic case it is difficult to obtain an explicit expression for $h_1$ and $h_2$, but in the large-$N$ limit we are interested in the harmonic function $H$ for the single-step flow of~\cref{eq:unsmeared_branes} can be written in terms of $\epsilon \equiv \frac{N_1 - N_2}{N_1} \ll 1$. For $\epsilon = 0$ one recovers the isotropic result, and thus by analyticity matching the structure of~\cref{eq:H_bergshoeff} to~\cref{eq:harmonic_function} via~\cref{eq:unsmeared_branes} shows that
\begin{eqaed}\label{eq:large-N_anisotropy}
    h_1 = h_2 + \epsilon^{\gamma} \, \delta h_1 + \dots
\end{eqaed}
for some $\gamma \geq 1$. Generically $\gamma = 1$, but it could be larger if the coefficient of the linear correction vanishes. Correspondingly, one also has $X_i^\text{max} \sim X_i^0 + \epsilon^\gamma \, \delta X_i$ for some $\gamma \geq 1$. Thus, the contribution of anisotropy to the single-step distance in~\cref{eq:anisotropy_upper_bound} cannot be parametrically larger than the contribution from the radion in~\cref{eq:radion_distance}, which also bounds the total single-step distance from below. All in all, the scaling of the discrete distance remains logarithmic, although computing the precise $\mathcal{O}(1)$ prefactor is more difficult in this less symmetric case. The above reasoning was kept quite general because it can be easily generalized to other distributions of branes, isolating the anisotropy contribution proportional to $\frac{N_1 - N_2}{N_1}$.

\section{Conclusions}\label{sec:conclusions}

We have introduced a notion of distance for discrete landscapes of vacua which takes into account their isolated nature. The idea is based on transitions between vacua mediated by domain walls (or vacuum bubbles in the metastable case).

While difficult to compute in general, it seems that many top-down settings arising from string compactifications afford simple bounds that are compatible with a logarithmic scaling. This pattern points to the idea of a discrete distance conjecture, whereby towers of species at large (discrete) distances become light exponentially fast with respect to this distance. On general grounds, obtaining bounds that are both sharp and useful is difficult due to how the distance is defined, but we have provided some estimates that may describe it qualitatively, at least in the context of BPS domain walls in supergravity.

Starting from a theory with a moduli space which satisfies the ordinary distance conjecture and lifting its moduli space to a discrete set of isolated vacua, one may wonder whether the persistence of exponential decay with respect to the discrete distance induced by domain walls is a swampland condition. If so, such a criterion would be verifiable entirely within the EFT landscape and applicable to EFTs with no moduli, both reasons which make this proposal attractive from a pragmatical perspective. It is conceivable that some EFTs satisfying the former distance conjecture but not the latter exist, but more exploration is needed to understand whether such theories, if any, are inconsistent, and why. Perhaps a first indication to an answer is that simple examples from string theory seem to satisfy this criterion due to hyperbolic field spaces with particular polynomial scalings, although one can construct toy models which also have this property.

To the extent that generalizing from conformal manifolds to RG flows is sensible, the holographic matching that we discussed in~\cref{sec:holography} is compelling and seems to support this view, and if further developed it can provide another tool to test proposals for holographic dual CFTs to gravitational EFTs with no known string theoretical construction. Hopefully, the first steps laid out in this paper can be useful to investigate these issues further.

\section*{Acknowledgements}

We thank Niccolò Cribiori, Stefano Lanza, Yixuan Li, Dieter L\"{u}st, Miguel Montero, Marco Scalisi, Irene Valenzuela, Vincent Van Hemelryck, Nicolò Petri, Thomas Van Riet and Flavio Tonioni for insightful discussions. We also thank Fotis Farakos, Vincent Van Hemelryck, Luca Martucci, Thomas Van Riet and Nicolò Risso for their feedback on the manuscript, and José Calder\'{o}n-Infante for collaboration in the initial stages of this project.

\printbibliography

\end{document}